\documentclass[aps,prb,preprint]{revtex4-2}

\usepackage{graphicx}
    \graphicspath{ {./figures/} }
\usepackage{amsmath}
\usepackage{amssymb}
\usepackage{array}

\begin{document}

\title{Labor-based grading practices in the physics classroom}

\author{Jeremy M. Wachter}
\email{wachterj@wit.edu}
\affiliation{School of Sciences \& Humanities, Wentworth Institute of Technology, Boston, MA 02155, USA}

\begin{abstract}

I describe the assessment framework of \emph{labor-based contract grading} (LBCG). In a labor-based grading scheme, the time and effort (``labor'')  a student spends on an assignment determines the credit they receive; the contract component requires students to design projects with clearly-defined goals and deliverables which must be satisfied to earn credit. LBCG is intended to promote student agency and engagement, and to provide a more equitable assessment framework given that students come with a wide range of prior experiences and preparation.  I illustrate the LBCG framework within the context of an upper level physics course, using a particular assignment as an example; I also provide information on student experiences and engagement.

\end{abstract}

\maketitle

\section{Introduction}\label{sec:intro}

Labor-based grading philosophy asserts that \emph{the student's labor creates value, and so the student's labor should determine their grade.} The word \emph{labor} is intended as a catch-all for the time and effort spent engaging with an assignment. Labor includes (but is not limited to) the following: solving problems, thinking about the class material, finding external resources, and explaining one's thought process, progress, and challenges. If a student provides evidence of such labor, they should receive credit for it. The regular practice of skills is one of the most important aspects of meaningful learning, \cite{doi:10.1080/00091380309604109,NAP9853} and so labor-based grading focuses on rewarding practice which is intentional, performed on different kinds of questions, and supported by substantive feedback.

It may seem extreme to ignore correctness in assessment,  however there is evidence suggesting that the pressures associated with being graded can reduce knowledge retention, motivation, and participation.\cite{retention,motivation,participation} Further, an undue focus on grades can impair academic performance or promote undesirable attitudes towards learning.\cite{impair,avoid,BTS} There are also equity concerns: within physics specifically, traditionally-graded assessments can negatively impact the achievement of women, under-represented minorities, and first-generation students~\cite{PhysRevPhysEducRes.16.020114,PhysRevPhysEducRes.16.020125,PhysRevPhysEducRes.18.020103} which alternative grading schemes can alleviate.\cite{PhysRevPhysEducRes.15.010132,richard2022implementing} In a labor-based grading scheme, students must still meet standards when presenting results and explaining their process, and the instructor remains active in assessing student work. Importantly, these standards and practices ensure that students still leave the course with an understanding of the material; this is discussed in the context of a specific sequence of assignments in Sec.~\ref{sec:example}.

This work discusses a scheme known as \emph{labor-based contract grading} (LBCG). The \emph{contract} component of this scheme asks students to design their own projects and explain not only what they will be doing, but also what proficiencies (skills, knowledges, etc.) they will demonstrate in completing the project. In this way, LBCG emphasizes student agency in the learning process (improving their engagement with and retention of the material \cite{doi:10.1080/09500693.2013.825066,PhysRevPhysEducRes.16.010109,PhysRevPhysEducRes.17.020128}).    Courses employing LBCG are more likely to meet the needs of a diverse group of learners, in keeping with research-backed approaches such as Universal Design for Learning (UDL). \cite{UDL}  The ``multiple means of expression principle'' of UDL states that curriculum design should allow learners alternative ways to demonstrate their knowledge, instead of always having to use the means selected by the instructor; LBCG aligns well with this principle. LBCG's philosophy and practices partially overlap with those of specification-based~\cite{SpecGrading,SpecGradingResponse} or competency-based~\cite{PhysRevPhysEducRes.13.020130,doi:10.1080/07377363.2016.1177704} grading, but key differences include 1) the students design the specifications/competencies (with instructor guidance), and 2) any scoring rubric is reduced to a simple binary of complete/incomplete. LBCG could also be considered a type of ungrading \cite{ungrading}, particularly in its emphasis on formative over summative feedback, although the method of assigning final grades discussed here is more concrete than most ungrading frameworks.

The LBCG framework was implemented in an upper-level course on mathematical and computational methods in physics, typically enrolling $\lesssim 10$ students, at Skidmore College, a small liberal arts college in the northeastern U.S. Prior instructors identified difficulties with the course and indicated that it could benefit from reform. These instructors commented on the difficulty of teaching such a wide-ranging course to a student body with extremely varied preparation. The programming experience of students in physics courses at Skidmore varies from absolutely none up to students who have done significant programming in research or upper-level mathematics or engineering courses.  The same variety of preparation also exists for mathematics---multiple students in my classroom were pursuing mathematics minors or second majors, and had prior exposure to topics like complex analysis or group theory.  The difficulty in designing traditionally-graded items that are both fair to the novice and challenging to the experienced student led me to consider a scheme where the student's effort determines their grade, rather than their prior experience. 

Classrooms with varied preparations are not unique to physics, nor is the desire for students to develop their own approaches.  In this regard, the LBCG philosophy can be applied to a wide variety of subjects (see \cite{LBCGExamples,OnlyLabor} for an example from biology).  Inoue \cite{Inoue} applied LBCG to a first-year writing course and their work is perhaps the most direct inspiration for the grading scheme discussed in this paper.

Inoue raised many concerns about traditional assessment, including: structural racism in traditional grading schemes (and academia more broadly); how assigning a singlular grade to a work often impedes meaningful discussion about that work; the emphasis on ``the correct way to write'' (which Inoue links to white supremacy) over students developing their own voice; and a personal dissatisfaction with the instructor experience of grading.  Largely, these points are transferable to other disciplines, physics included.  At first sight, one might think there is always a ``correct'' answer in physics, but in fact physics students must engage in modeling physical systems and designing experiments.  Activities such as these have no ``correct'' answer; they require students to develop their own approaches and find their (physics) voice.

While this work discusses labor-based assessment for a particular course, the essential ideas (of mindful practice, student freedom, and decreased biases and pressures associated with traditional grading) can easily be applied to other courses. However, it is worth mentioning three significant features of the course.
\begin{itemize}
    \item \underline{The class sizes were small.} Both iterations of the course had only eight students enrolled. Providing formative feedback and managing extra assignments can be time- and labor-intensive and can be more  manageable in a small class.
    \item \underline{This was an upper-level course for physics/STEM majors.} It is likely that the students were more motivated to learn than in introductory courses. I don't think it's impossible to apply these ideas to an introductory course, but the differing motivations, interests, and levels of engagement of students in those courses could make the implementation of LBCG more difficult, and significant adjustments might be required.
    \item \underline{There was significant in-class discussion and small-group work.} The course used a flipped classroom approach. This provided frequent opportunities for informal formative assessment, where students received instant feedback on their initial problem-solving attempts. This was especially important for students learning to program for the first time.
\end{itemize}

\section{Course outline}\label{sec:ul}

The course described in this paper is a 13 week-long, required upper-level course for physics majors. Its pre-requisites are modern physics and linear algebra. Most students were physics majors in their third or fourth years, although other STEM majors or second-year physics students would occasionally enroll. I am reporting on two iterations of this class: the first iteration in Spring `21 was hybrid (half in-person and half online), and the second iteration in Spring `22 was fully in-person.

Both times the course was offered, it began with an introductory module on programming; additional programming techniques and concepts were introduced and reinforced throughout the semester.\footnote{In the first iteration, all students did the same readings and worked on the same programming materials, with more challenging questions available to students who wished to pursue them. In the second iteration, the more experienced students were given a parallel track to learn about more advanced programming topics, although there was still cross-collaboration in-class.} On a typical day, students reviewed and discussed the day's readings during the first half of class, and the second half of class was devoted to small-group work.  Every student began the course with a B, which could then be changed by the completion or non-completion of the two forms of assessed items: core assignments, and extra assignments. The most salient features of these assignments are given below, with additional details in Appendix~\ref{app:assignment-details}.

Core assignments were weekly problem sets which contained four to six ``short prompts'' (similar to in-class exercises) and one or two (multi-part) ``long prompts''. \footnote{Students were expressly permitted and encouraged to work collaboratively on core assignments, with the requirement that they cite everyone they worked with in their final submission.} The core assignments were counted as complete if two criteria were met:
\begin{enumerate}
	\item The student demonstrated documented effort on all problems on the assignment.
	\item The assignment was turned in on time.
\end{enumerate}
\emph{Documented effort} means providing a comprehensive record of work, but does not require getting the correct answer or even reaching a final result. In this way, even if a student doesn't complete a problem (in the traditional sense), they still benefit from having to explain their reasoning and justify their process. It is their \emph{labor} which is given credit. This criterion was discussed on the first day of class and defined in the syllabus; precise wording can be found in Appendix~\ref{app:core-details}.

If a core assignment was recorded as complete, the student's grade was unchanged. If it was recorded as incomplete, the student's grade was decremented by one step (e.g., B+ to B). Because the labor-based grading scheme was unfamiliar to students, flexibility was sometimes needed while students adjusted to the approach.  For example: each semester, some students neglected to turn an assignment they had worked on because they thought the assignment was ``incomplete". This habituation to traditional grading was remedied by a follow-up conversation and clarification of the LBCG scheme.

The second type of assessed item was the extra assignment, which employed the \emph{contract} part of \emph{labor-based contract grading}. These assignments were entirely student-designed, and the lifecycle of an extra assignment looks something like the following. The student approaches me with an idea for a project, which we discuss. The student then drafts a contract which clearly outlines the project, deliverables, and proficiencies to be demonstrated. After some iteration, the contract is agreed upon and the student begins work. Once the student is satisfied with that work, they submit the project materials; I review them and, if necessary, request revisions required in order to meet the contract conditions and for the assignment to be considered complete.\footnote{The syllabus included a clause that students could argue for why some work which I had marked for revision did, in fact, satisfy the proficiencies. In practice, students just made the changes and additions I requested.} Two examples of extra assignment contracts, along with two additional extra assignment topics, can be found in Appendix~\ref{app:sample-contracts}.

Once an extra assignment was recorded as complete, it incremented the student's grade by one step (e.g., B to B+). Incomplete extra assignments could never decrement a student's grade. In the first iteration of the course, some students had difficulty finding the time to complete extra assignments at the end of the semester, which I discuss further in Sec.~\ref{sec:completion} and the supplemental material.\footnote{Reflections on my experience as an instructor, as well as a discussion of potential changes to my implementation of LBCG, can be found at [url to be inserted by AIPP].}

In this scheme, it is refreshingly simple to calculate a student's final grade. From the starting grade of B, each incomplete core assignment decrements the grade one step, and each complete extra assignment increments the grade by one step. A student therefore knows their grade at any time with near certainty. I chose B to be the starting grade based on two principal considerations. First, I intended to make the core assignments involved and important to student learning; completing all of them on time would leave the students with a good understanding of the material and would represent a `B' effort by the standards of the institution. Second, I wanted to make high grades reasonably accessible to students---a student needs to complete three extra assignments to reach an A. The final grade distributions were comparable to other upper-level physics classes at the institution.

\section{Labor-based assignments in practice}\label{sec:example}

Since they account for a significant portion the final grade, it's important to discuss how core assignments alone meet the course learning goals. I discuss how programming and physics knowledge are developed via the LBCG approach by looking at two long prompts on the discrete Fourier transform (DFT) from two sequential core assignments. 

As `long prompts', each included at least a page of background information. They are summarized as follows:
\begin{enumerate}
    \item The students are introduced to signal analysis and the DFT, including example applications in physics and engineering as well as to relevant terminology (bandwidth, sampling frequency, time series, etc.).  They are then given an audio file of a beat-frequency signal, along with a mathematical description of the signal, and asked to compare the DFT of the audio file to the continuous Fourier transform (CFT) of the mathematical description. This last step required them to write their own ``na\"ive'' version of the DFT (an implementation which is \emph{not} a fast Fourier transform). 
    \item The students are given three different sound files: one is a computer-generated 440 \,Hz (A4) pure tone, one is a computer-generated 440\,Hz tone with five overtones, and one is a recording of an actual instrument playing the same note. The students are then given the definition of a signal's power spectrum and are then asked to determine appropriate sample sizes for analyzing the sound files. Finally, they are asked to produce plots of the power spectra for all three sound files and comment on the differences.
\end{enumerate}

The students are asked to use the analysis and code they wrote for the first prompt as a basis for working on the second prompt. A student who has non-functioning code or a gap in their understanding of the DFT after completing the first prompt, must revisit that submission (and the feedback I provided) in order to make meaningful progress on the second prompt. In practice, a majority of students had a working (though not necessarily free of bugs) DFT after working on the first prompt, and all students had a working (same caveat) DFT after the second prompt.  Most students gave a satisfactory analysis of the sample size, sampling rate, etc. for the sound files and the DFTs they performed. All students made clear connections between the dominant peaks of the power spectrum, the constituent frequencies, and the perceived sounds.

I used a similar structure in a module covering numerical solutions to differential equations, where students built up a fourth-order Runge--Kutta solver starting from a forward-difference method applied to simple systems which could be solved analytically (exponential decay, damped harmonic oscillation), and culminating in a (two-dimensional) three-body problem. Again, many students were able to get a working RK4 solver and all students were able to attempt the three-body problem with at least a lower-order solver.

There are two major factors here which are important for the success of the LBCG scheme. First, sequences of related prompts were used---questions on core assignments relied on ideas, code, or outcomes from prior assignments. These are crucial because if the students are not tested on prior material, or are not in some way asked to revisit and revise earlier work, they can leave with an incomplete understanding.  Second, detailed, formative feedback is provided on the core assignments. \footnote{By \emph{formative}, I mean feedback which doesn't focus on correct/incorrect, but which is more in the spirit of comments on a draft. I typically highlighted student work which showed good thinking, asked them to explain or elaborate upon points which were weak, pointed out contradictions or inconsistencies, and suggested resources which might be helpful based on the misconceptions or gaps in understanding I saw. For small syntax or typographical errors in code, I typically just told the students the fix.} Students who didn't correctly answer a prior prompt need sufficient commentary to productively build upon that work. For the instructor, this is probably the most labor-intensive part of a LBCG course, since substantial comments must quickly be written so that they're of use to the students on future core assignments. 

Implemented this way, we might understand the labor-based approach as a species of formative assessment, wherein the students are not penalized for developing their understanding over a longer timeframe, and addressing initial misconceptions is treated as a natural part of the learning process.\cite{misconceptions} However, in contrast to other schemes (which might use formative assessment as a supplement, or which might additionally employ summative assessments), here the \emph{entirety} of a student's grade comes from them putting in the work and, optionally, meeting the standards they choose for themselves.

\section{Assessment completion rates}\label{sec:completion}

One way to evaluate the effectiveness of the LBCG scheme is by measuring assignment completions (see Table~\ref{tbl:core-extra}).

\begin{table}
    \begin{tabular}{lr|c|c}
    	                         &           & Iteration 1 & Iteration 2\\
        \hline
        Students:                &           & 8           & 8\\\hline
        Contracts approved:      & Total     & 15          & 17\\
                                 & Median    & 2           & 3\\
                                 & Range     & $[0,4]$     & $[0,3]$\\\hline
        Contracts completed:     & Total     & 13          & 16\\
                                 & Median    & 1.5         & 2.5\\
                                 & Range     & $[0,4]$     & $[0,3]$\\\hline
        Core assignments missed: & Total     & 0           & 4\\
                                 & Median    & 0           & 0\\
                                 & Range     & $[0,0]$     & $[0,3]$
    \end{tabular}
    \caption{The number of contracts (extra assignments) approved and completed, as well as core assignments missed, for both iterations of the course. For multi-student contracts (assignments completed in a team), \emph{each} student in the group is included in the ``Total''. All students remained enrolled for the entirety of the course in both iterations.}\label{tbl:core-extra}
\end{table}

The number of missing core assignments is notably different between the two iterations. Most of that difference is explained by one student in the second iteration, who missed three core assignments. 

Student engagement with the extra assignments was similar in both iterations, with better overall engagement in the second iteration. In both iterations, two students elected not to attempt any extra assignments. \footnote{I spoke to them and ensured they understood what that meant for their grade. One student said that they were happy to have a ``guaranteed'' B if they were conscientious about completing all of the core assignments, and preferred to focus additional effort on other classes; another said that they were fine with a B and were satisfied with their understanding of the material from working only on the core assignments.}

\begin{figure}
	\centering
	\includegraphics[scale=0.80]{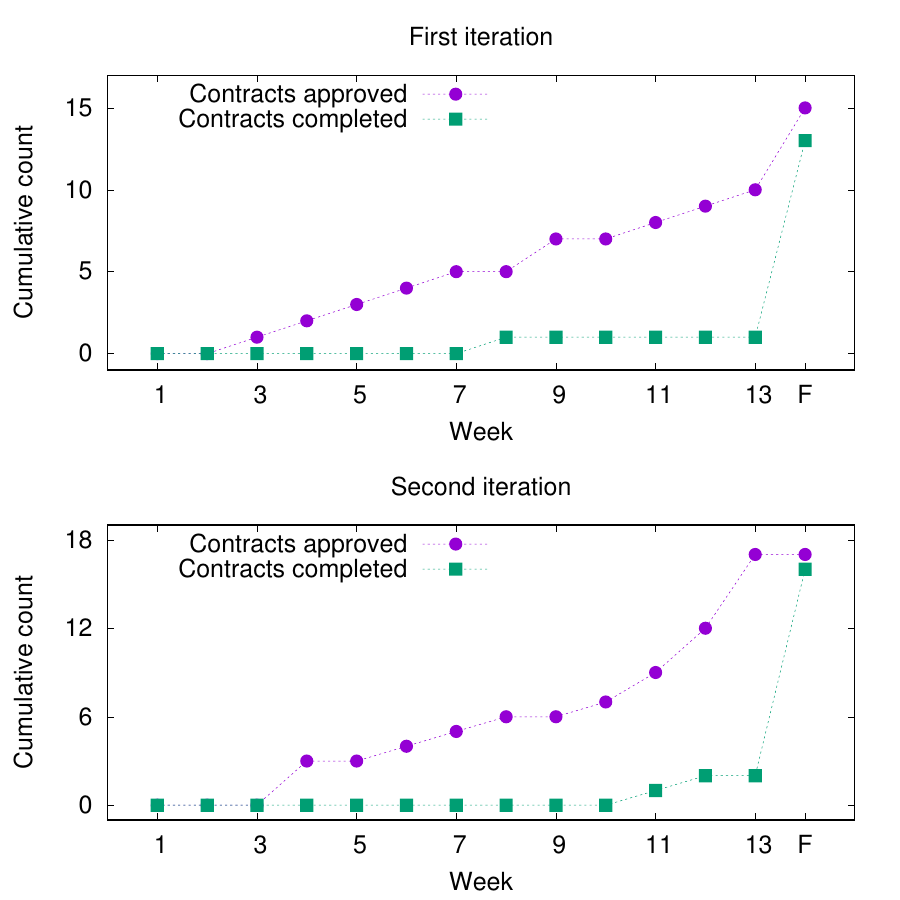}
	\caption{The total contracts approved (circles) and completed (squares) in each week of the first (upper panel) and second (lower panel) iterations of the course. The ``F'' indicates ``study days plus finals week'', as students were allowed to submit completed extra assignments until the end of the testing period. For multi-student contracts, the count was incremented by the number of students involved.}\label{fig:contracts-t}
\end{figure}

The total number of proposed and completed contracts, by week, is shown in Fig.~\ref{fig:contracts-t}. This illustrates one potential concern: students might attempt multiple contracts at the end of the semester, leaving themselves a large workload and increasing the chance that they wouldn't complete a contract on time. 

In the first iteration, the deadline on contract submission was the first day of finals week. Students were also aware that the review process could take a few days, and that they should anticipate at least one round of revisions.  Still, the majority of contract completions took place during the end-of-semester ``study days plus finals week'' period. In response, I modified the rules for the second iteration to prohibit proposals after the end of classes. While most contract completions still took place at the end of the semester, the change reduced the time crunch for both the students and me. This change is also visible in the nature of the incomplete contracts: only one contract in the second iteration was abandoned during the study period so that the student could focus on other courses, while in the first iteration, both incomplete contracts were due to students not having time to revise incomplete work.

\section{Student feedback on the labor-based contract grading scheme}\label{sec:ul-feedback}

End-of-semester questionnaires asked students to reflect on their own level of commitment to the course, and to comment on the course. The questionnaire primed students to comment in the context of course learning goals and to focus on specific examples when answering.

Below, I provide a selection of student comments specific to the labor-based grading scheme, taken from both iterations of the course. The response rate was 50\% in the first iteration (the hybrid class; response rates suffered college-wide that semester) and 87.5\% in the second iteration. Roughly half of the respondents specifically commented on the labor-based grading scheme in both iterations.

\begin{quotation}
    ``I also really enjoyed the grading system. It really took the stress off homework when learning new/weird concepts \& made completing homework much more enjoyable.''
\end{quotation}

\begin{quotation}
    ``I like the grading policy. The assignments were a bit challenging esp. for students who have none/a little coding background. So this policy helped a lot!''
\end{quotation}

\begin{quotation}
    ``The grading scheme saved me a serious load of stress this semester. The focus on completing assignments vs. getting answers correct is one of my favorite aspects of the physics department.''
\end{quotation}

\begin{quotation}
    ``Great class, the grading scheme is very conducive to learning and challenging oneself.''
\end{quotation}

This suggests that the LBCG scheme was well-received both by students who indicated less comfort with the material as well as by students who were hoping for a more demanding workload. While other comments related to the LBCG scheme have been omitted, none of those comments were critical of it. Students did criticize of other aspects of the class in their questionnaire responses, such as the textbook being too dense or the pace of course being too fast.

\section{In closing}\label{sec:end}

The labor-based contract grading scheme was successful when measured in terms of student engagement, student feedback, and instructor experience. When it comes to learning outcomes and knowledge retention, it's more difficult to directly compare the LBCG scheme with a more traditional scheme, as I lack both an appropriate comparison course and a properly designed assessment instrument for this course. 

Overall, I found the scheme very well-suited to my pedagogical goals. The increase in student agency (and thus involvement) and the emphasis on improvement through doing the work are, to me, the two greatest advantages of this approach to assessment. Readers may have motivations similar to mine or not; regardless, I would encourage readers to consider implementing a labor-based grading scheme in their own classes.

\section*{Acknowledgements}

I would like to thank Evan Halstead for helpful comments and discussion on a draft of this work, Kendrah Murphy for useful conversations and feedback after a class visit, and Greg Gerbi for advice on teaching the course and for providing materials. I would also like to thank the reviewers for suggesting additional references as well as making various suggestions for strengthing the argument for LBCG's utility.

\section*{Author Declarations}

I have no conflicts to disclose.

\appendix

\section{Details on core and extra assignment implementation}\label{app:assignment-details}

\subsection{Core assignments}\label{app:core-details}

\subsubsection{Documented effort}

The definition in the syllabus was:
\begin{quotation}
    The term \emph{documented effort} means that you have written down, or in some other way made a record of, your process for solving (or attempting to solve) a problem. It is not sufficient to just write down some equations which you think might be useful, or to repeat the setup given in the problem prompt; I am looking for evidence that you have engaged with the material, even if you didn't reach a final conclusion. Likewise, it is not sufficient to say that you thought about the problem and couldn't reach a conclusion; you should be as specific as possible with what you thought about, making reference to what parts of the text or other resources you looked for, what related problems you might have looked to for inspiration, or what discussions you may have had with classmates in attempting to solve a problem. If you're ever uncertain about whether or not what you've done satisfies the threshold of documented effort, you can always ask me!
\end{quotation}

\subsubsection{The freebie rule}

All students were given one ``freebie'', that is, one core assignment which they did not have to turn in and which did not count against their grade. The intent of this rule was to account for busy periods of the semester, extracurriculars, or other external demands on the student's time which might prevent them from completing a given core assignment. It was popular for students to use the freebie on the final core assignment of the semester; this was foreseen, and given how busy the end of the semester typically gets, I consider it a prudent usage.\footnote{Here is a noteworthy departure from Ref.~\onlinecite{Inoue}: I did not allow students to convert an unused freebie into an increment to their grade at the end of the semester. This is because such a policy, in effect, just makes the baseline grade one increment higher. Using the freebie doesn't really prevent a decrement in grade; it just means that the student is giving up a later increase in the grade, and thus are still materially penalized for using the freebie.} I did require that students inform me that they were using the freebie before the assignment was due.

The definition in the syllabus was:
\begin{quotation}
	\textbf{Freebie rule:} every student gets one freebie for the semester. This means that for any core assignment, you can invoke your freebie to have the missing assignment not count against your grade. You do not need to give any reason or justification when you invoke the freebie rule; \emph{the only condition} for the freebie rule is that you must invoke it before the assignment is due.
\end{quotation}

\subsection{Extra assignments}\label{app:extra-details}

Here I give a more detailed description of the extra assignment process, from conception to completion, as well as more information about how extra assignments were introduced and managed.
\begin{enumerate}
	\item The student approaches me with an idea for a project related to the course topics. We discuss what topics are relevant, what proficiencies might be demonstrated, and what form the final deliverable will take. The topic and the form of the deliverable were unrestricted, so long as the student could meaningfully connect it to some pertinent mathematics or physics. 
	\item The student writes a contract that includes a description of the project, the final deliverable (including length/size/complexity estimates), and what proficiencies the student will demonstrate in undertaking this work. The satisfaction of these proficiencies determines when the assignment is complete.
	\item I read the contract and make suggestions; it undergoes some number of revisions until both the student and I are satisfied, at which point the contract is signed and the assignment is considered underway.
	\item The student submits the final deliverable. I review the work and, if necessary, request modifications or clarifications to satisfy the proficiencies. After some number of revisions, I approve the deliverable as satisfying the proficiencies laid out in the contract, and the extra assignment is recorded as complete.
\end{enumerate}
While I refer here to ``the student'', I did permit multi-student projects, provided the division of labor (and thus who would demonstrate which proficiencies) was clear, and the deliverable's length/size/complexity was commensurate with the work of multiple students.

I provided the students with \LaTeX~templates for contracts, with very broad language and several fill-in-the-blanks parts, to give them an idea of what the proposal should look like, as well as what a ``proficiency'' might be. I also told them that I expected each extra assignment to take them about as long as a core assignment to complete in terms of total time spent working.

Students could have any number of extra assignments underway at any given time; although I could always deny a new contract's approval until a student completed some of their existing assignments, in practice I never encountered this problem. I did not accept new proposals after a particular date late in the semester, and all extra assignments were due by the end of finals week, but I did not otherwise impose time constraints or ask the students to include a timetable in their contracts.

\section{Sample contracts and projects}\label{app:sample-contracts}

In this section, I first provide two examples of student-proposed extra assignment topics followed by two examples of student contracts.

In the first iteration, a student was interested by the difference in efficiency between the ``na\"ive'' DFT approach (with computational complexity $\mathcal{O}(n^2)$) and the fast Fourier transform used in real applications($\mathcal{O}(n\log n)$); they investigated the origin of this difference, wrote their own implementation of the Cooley--Tukey algorithm, \cite{CooleyTukey} and tested both versions to verify the relationship. In the second iteration, a student examined the spectra of various guitar effects pedals and explained how the changes in those spectra relate to changes in sound. Both students needed to have a sufficiently-developed conceptual understanding of the DFT in order to conceive of, propose, and complete these projects.

Below, I provide two reproductions of student-designed contracts for extra assignments to give the reader a sense of what they might look like. Both have been lightly edited to preserve student anonymity, and both resulted in a completed extra assignment. The first is a contract for a presentation on string worldsheets, and was prepared by a sophomore in the first iteration of the course. (Several students elected to do presentations on some topic, which were delivered to the entire class during finals period). The second is an experimental project which involves recording various electric guitar signals and performing frequency spectrum analysis to describe differences in sound, and was prepared by a senior in the second interation of the course. All contract revisions and all feedback regarding the projects were done via email to ensure a written record.

The similar language in both contracts, ``clear and accessible to an audience of my peers'' is my suggested language from one of the contract templates. Its purpose is to focus the student's consideration of how they will communicate information as well as what information should be explained versus considered background knowledge. A discussion about how well that was achieved and how it might be improved was always a part of the feedback process.

\subsection{Contract for ``Applications of Complex Analysis to String World-Sheets''}\label{app:string-theory}

\subsubsection*{Description of work to be undertaken}

To complete this contract, I will give a presentation on the applications of complex analysis to the treatment of string world-sheets as Riemann surfaces. This presentation will be 12 minutes long, with 3 minutes for questions, with a general structure as follows:
\begin{itemize}
	\item Presentation of relevant background topics/concepts, to include, e.g., definitions of a world-sheet, Riemann surface, and string. In presenting these topics, I will also provide explanations of concepts prior to them. These definitions will not be comprehensive, but should be sufficient for obtaining a complete understanding in the context of our discussion.
    \item Connection between these topics and our class discussions of the Cauchy-Riemann equations, which will lead us into the next item:
    \item Mathematical and conceptual exploration of the world-sheet for a free, open string
\end{itemize}
If I include a figure or illustration, I will either create my own or direct the audience to the appropriate figure in a referenced work. Most equations will probably not be my own derivations---in these cases, I will direct the reader to the original source. If I show a derivation or special case I worked out myself, I will also make this clear to my audience.

\subsubsection*{Proficiencies to be demonstrated}

The presentation will demonstrate the following proficiencies:
\begin{itemize}
	\item Conceptual fluency: I will demonstrate my understanding of the use of complex analysis to examine strings' history in spacetime by presenting illustrative calculations with conceptual support. All explanations will be in my own words, to a level which leaves the audience with a sense of being introduced to the idea of treating string world-sheets as Riemann surfaces.
	\item Scientific communication: The presentation will be clear and accessible to an audience of my peers.To achieve this, I will give definitions of all technical terms not covered in class!
	\item Literature review: I will include a bibliography containing all sources from which I obtained the information presented. It is likely that my chief source will be Barton Zwiebach's textbook, ``A First Course in String Theory''.
\end{itemize}

\subsection{Contract for ``Guitar Signal Construction Using Fourier Series Analysis''}\label{app:guitar}

\subsubsection*{Description of Project}

To complete this contract, I will record different electric guitar signals through different effects pedals and use the discrete Fourier transform to decompose them into frequency spectra to describe the audible effect introduced by each pedal. The project will consist of the following deliverables:
\begin{itemize}
    \item Multiple recordings of a single guitar note subject to multiple different effects: clean(control), fuzz, and rotor, using Audacity.
    \item Raw data files associated with the above.
    \item A Python script used for processing data.
    \item A report discussing:
    \begin{itemize}
        \item Brief introduction to effects pedals.\
       \item Experimental procedure detailing the recording process for each signal and a description of each effects pedal/device used in doing so.
       \item Raw signal data and frequency spectra for each signal using \texttt{matplotlib.pyplot} and \texttt{numpy.fft.fft()} in Python.
       \item Discussion on how data/frequency spectra relates to how each pedal modifies the output signal of a guitar.
    \end{itemize}
\end{itemize}

\subsubsection*{Proficiencies to be demonstrated}

The completed project will demonstrate the following proficiencies:
\begin{itemize}
    \item Experimental implementation of discrete Fourier transforms by using them in a practical application. I will explain the experimental process to a level which leaves the reader with an introductory idea of effects pedals, and how discrete Fourier transforms can be used to analyze them, without requiring specialist knowledge or detailed study.
    \item Scientific communication: The resulting paper will be clear and accessible to an audience of my peers.
    \item Scientific figure creation: I will create raw data plots, and use the discrete Fourier transform to construct the frequency spectra of different guitar signals to support my paper, using the \texttt{matplotlib.pyplot} package to do so. I will include the code used to generate the figures, as well as any data files, in my final submission.
\end{itemize}

%
%

\end{document}